\documentclass[aps,prb,onecolumn,preprint,showpacs,superscriptaddress]{revtex4}
\usepackage{graphicx}
\usepackage{dcolumn}
\usepackage{bm}
\usepackage{amssymb}
\usepackage{color}

\begin{document}
\title{Biharmonic Split Ring Resonator Metamaterial: Artificially dispersive effective density in thin
periodically perforated plates}
\author{Mohamed Farhat}
\affiliation{Institute of Condensed Matter Theory and Solid State Optics, Abbe Center of Photonics, Friedrich-Schiller-Universit\" at Jena, D-07743 Jena, Germany}
\affiliation{Division of Computer, Electrical, and Mathematical Sciences and Engineering, 4700 King Abdullah University of Science and Technology, Thuwal 23955-6900, Saudi Arabia}
\author{Stefan Enoch}
\affiliation{Aix-Marseille Universit\'e, CNRS, Centrale Marseille, Institut Fresnel, \\
Campus universitaire de Saint-J\'er\^ome, 13013 Marseille, France}
\author{Sebastien Guenneau}
\affiliation{Aix-Marseille Universit\'e, CNRS, Centrale Marseille, Institut Fresnel, \\
Campus universitaire de Saint-J\'er\^ome, 13013 Marseille, France}

\date{\today}

\begin{abstract}
We present in this paper a theoretical and numerical analysis of bending waves localized on the boundary of a platonic crystal whose building blocks are split ring resonators (SRR). We first derive the homogenized parameters of the structured plate using a three-scale asymptotic expansion in the linearized biharmonic equation. In the limit when the wavelength of the bending wave is much larger than the typical heterogeneity size of the platonic crystal, we show that it behaves as an artificial plate with an anisotropic effective Young modulus and a dispersive effective mass density. We then analyze dispersion diagrams associated with bending waves propagating within an infinite array of SRR, for which eigen-solutions are sought in the form of Floquet-Bloch waves. We finally demonstrate that this structure displays the hallmarks of All-Angle-Negative-Refraction(AANR) and it leads to superlensing and ultrarefraction effects, interpreted thanks to our homogenization model as a consequence of negative and vanishing effective density, respectively.

\end{abstract}

\maketitle

\section{Introduction}

Left Handed Materials (LHM) are a new kind of materials which were theoretically envisionned by Veselago \cite{veselago} as early as in 1967. Such materials have simultaneously negative relative permittivity ($\varepsilon_{\mathrm{r}}$) and negative relative permeability ($\mu_{\mathrm{r}}$). This theoretical curiosity became a real field of research in 2000 after Pendry showed the potential of LHM to overcome the diffraction limit \cite{pendry_prl00} and Smith \textit{et al.} \cite{smith_nr} proposed a first realization for such extra-ordinary materials based on periodic lattices combining Split Ring Resonators (SRR: Concentric annular rings with splits) and wires. The latter work can be considered as the experimental foundation of LHM (as the first experimental evidence of negative refraction) and it is based on a theoretical study by Pendry \textit{et al.} which shows that negative permittivity could be obtained by a periodic arrangement of parallel wires and that a periodic lattice of SRR had a negative magnetic response around its resonance frequency \cite{pendry_ieee99}.

\noindent In the recent years, there has been a keen interest in wave propagation in periodically structured media \cite{wavesPC}. Investigation of photonic crystals has paved the way to the theoretical prediction and experimental realization of photonic band gaps \cite{yablono87,john,krauss,zengerle87,gralak2000,notomi2002,luo2002} i.e. ranges of frequencies for which light, or a light polarization, is disallowed to propagate. Soon after, the focus has been extended to the study of acoustic waves in periodic media, and the existence of phononic band gaps has been verified both theoretically and experimentally \cite{2-ross1,2-ross2,acoustic,zhang2004,hu2004,feng2006}. Recently, the interest was even extended to the study of different types of waves, e.g. liquid surface waves \cite{chou97,chou98,mciver2000,torres2000,hu2003} or biharmonic waves \cite{ross-platonic,sasha-prsa,farhat-apl,farhat-epl,farhat-prl2009} in perforated thin plates. It has been shown that complete bandgaps also exist for these waves when propagating through a periodic lattice of vertically standing rods or over a periodically perforated thin plate \cite{torres2000,farhat-apl}. In addition, many interesting phenomena have been reported, including negative refraction  \cite{veselago,pendry2000,maystre2004,guenneau2005,smith2000,ramak2005}, the superlensing effect and cloaking \cite{schurig,norris,farhatprl,farhatwm, farhatpre}. The essential condition for the AANR effect is that the constant frequency surfaces (EFS: equifrequency surfaces) should become convex everywhere about some point in the reciprocal space, and the size of these EFS should shrink with increasing frequency \cite{zengerle87,gralak2000,notomi2002}.\\

\noindent In this paper, we focus on the application of split ring resonator (SRR) structures \cite{pendry_ieee99,Seb_SRR,guenneau_physb06} to the domain of elastic waves. We first derive the homogenized governing equations of bending waves propagating within a thin-plate with a doubly periodic square array of freely vibrating holes shaped as SRR, from the generalized biharmonic equation, and an asymptotic analysis involving three scales (one for the thickness of the thin-cut of each SRR, one for the array-pitch, and one for the wave wavelength). We then present an analysis of dispersion curves. To do this, we set the spectral problem for the biharmonic operator within a doubly periodic square array of SRR, homogeneous stress-free boundary conditions are prescribed on the contour of each resonator and the standard Floquet-Bloch conditions are set on the boundary of an elementary cell of the periodic structure. Such a structure presents an elastic bandgap at low frequencies. It turns out that the asymptotic analysis of our structure allows us to get analytically the frequency of the first localized mode and then the frequency of the first band gap.
The aim of our work is actually to demonstrate the AANR effect at low frequencies for elastic thin perforated plates as well as their superlensing properties. Ultra-refraction is also considered and shows the versatility and power of using such structured media to realize new functionalities for surface elastic waves.

\section{Homogenization of a thin-plate with an array of stress-free SRR inclusions near resonance}

The equations for bending of plates can be found in many textbooks \cite{timoshenko,graff}.
The wavelength $\lambda$ is supposed to be large enough compared to the thickness
 of the plate $H$ and small compared to its in-plane dimension $L$, i.e. $H \ll \lambda \ll L$.
In this case we can adopt the hypothesis of the theory of Von-Karman
\cite{timoshenko,graff}. In this way, the mathematical setup is essentially two-dimensional,
the thickness $H$ of the plate appearing simply as a parameter in the governing equation.

We would like to homogenize a periodically structured thin-plate involving resonant elements. The resonances are associated with fast-oscillating displacement fields in thin-bridges of perforations shaped as split ring resonators (SRR), and we filter these oscillations by introducing a third scale in the usual two-scale expansion. We start with the Kirchhoff-Love equation and we consider an open bounded region $\Omega_f\in\mathbb{R}^2$. This region is e.g. a slab lens consisting of a square array of SRR shaped as the letter $C$.

\begin{figure}[]
\begin{center}
\scalebox{0.6}{\includegraphics[angle=0]{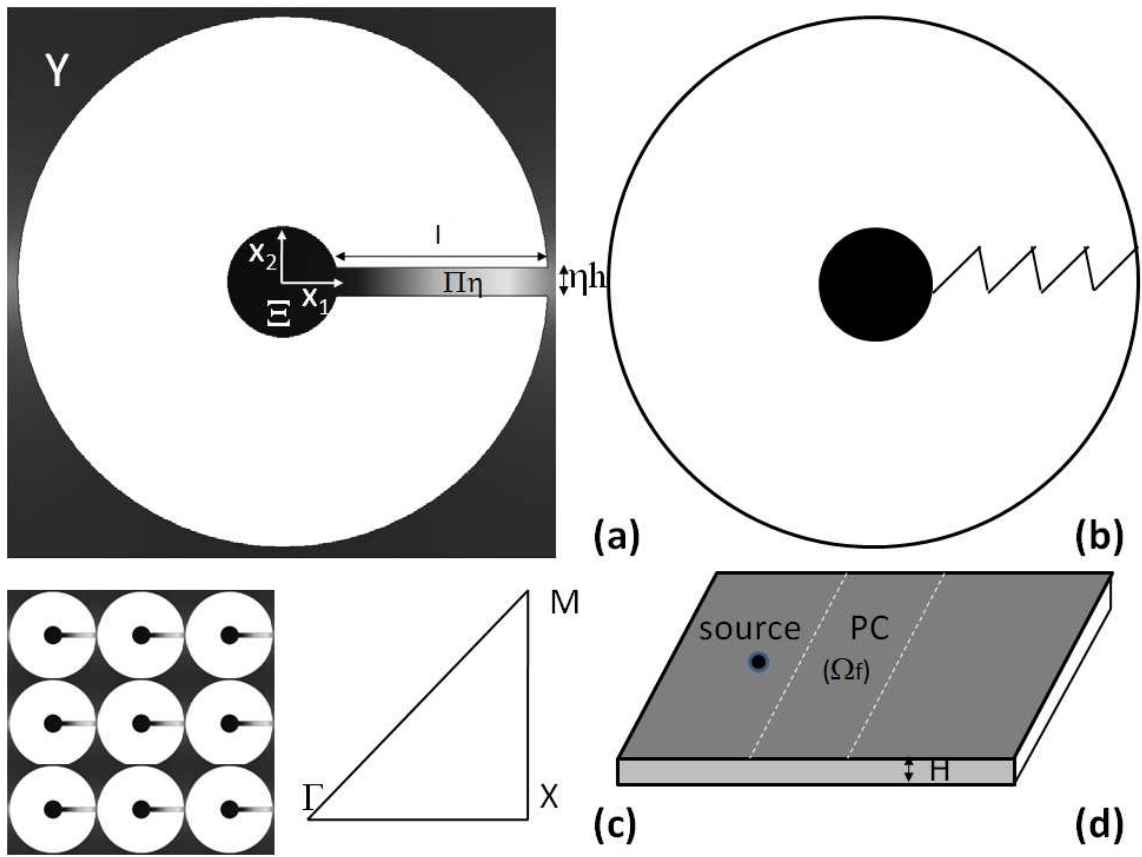}}
\caption{(a) Geometry of a split ring resonator C consisting of a disc $\Sigma$ connected to a thin ligament $\Pi_\eta$ of length $l$ and thickness $\eta h$ where $0<\eta\ll 1$ in a unit cell Y; (b) Helmholtz resonator consisting of a mass connected to a wall via a spring which models resonances of SRR in (a); (c) Doubly periodic square lattice of SRR with the first Brillouin zone $\Gamma$XM
in reciprocal space; (d) Geometry of the thin plate of thickness $H$, with a source on the left side of a platonic crystal (PC) slab occupying
the region $\Omega_f$.}
\label{fig6}
\end{center}
\end{figure}

\noindent When the bending wave penetrates the structured area $\Omega_f$ of the plate
whose geometry is shown in Figs. \ref{fig6}(c)-(d), it
undergoes fast periodic oscillations. To filter these oscillations,
we consider an asymptotic expansion of the associated vertical
displacement $U_\eta$ solution of the biharmonic equation given in (\ref{bihaeta}) in terms of a
macroscopic (or slow) variable ${\bf x}=(x_1,x_2)$
and a microscopic (or fast) variable ${\bf
x}_\eta={\bf x}/\eta$, where $\eta$ is a small positive
real parameter.

\noindent With all the above assumptions, the out-of-plane
displacement ${\bf u}_\eta=(0,0,U_\eta(x_1,x_2))$ in the
$x_3$-direction (along the vertical axis) is solution of
(assuming a time-harmonic dependence $\exp(-i\omega t)$
with $\omega$ the angular wave frequency):
\begin{equation}
\begin{array}{ll}
\displaystyle{\frac{\partial^2}{\partial x_1\partial x_1}\left({{D}_\eta}\left(\frac{\partial^2}{\partial x_1\partial x_1}+\nu_\eta \frac{\partial^2}{\partial x_2\partial x_2}\right)\right)U_\eta} \nonumber \\
+\displaystyle{\frac{\partial^2}{\partial x_2\partial x_2}\left({{D}_\eta}\left(\frac{\partial^2}{\partial x_2\partial x_2}+\nu_\eta \frac{\partial^2}{\partial x_1\partial x_1}\right)\right)U_\eta} \nonumber \\
+\displaystyle{2\frac{\partial^2}{\partial x_1\partial x_2}\left({D}_{\eta}\left(1-\nu_\eta\right)\frac{\partial^2}{\partial x_1\partial x_2}\right)U_\eta
-\,\beta_\eta^4\,U_\eta}
=0 \; ,
\end{array}
\label{bihaeta}
\end{equation}
inside the heterogeneous isotropic region $\Omega_f$ (the platonic crystal, PC, in Fig. \ref{fig6}(d)), where
$$D_\eta=D(\frac{{\bf x}}{\eta}) \; , \; \nu_\eta=\nu(\frac{{\bf x}}{\eta}) \; \hbox{ and }
\; \beta^4_\eta=\beta_0^4\rho(\frac{{\bf x}}{\eta}) \; , $$
are nondimensionalized spatially varying parameters related to the flexural rigidity of the plate, its Poisson ratio and the
wave frequency, respectively. In most cases, $D$ and $\nu$ take piecewise constant values, with
$D>0$ and $-1/2<\nu<1/2$. Note that $\beta_0^2=\omega\sqrt{\,\rho_0 H/D_0}$, where $D_0$ is the
flexural rigidity of the plate outside the platonic crystal, $\rho_0$ its density
and H its thickness.

Remark that (\ref{bihaeta}) is written in weak form and we notably retrieve the classical boundary conditions for a
homogeneous plate with stress-free inclusions (vanishing of bending moments and shearing stress for vanishing
$D_\eta$ and $\nu_\eta$ in the soft phase) \cite{graff}.
Since there is only one phase in the problem which we consider (homogeneous medium outside freely-vibrating inclusions),
it is also possible to recast  (\ref{bihaeta}) as
\begin{equation}
(\sqrt{D_\eta}\nabla^2 U_\eta+\beta_\eta^2)(\sqrt{D_\eta}\nabla^2 U_\eta-\beta_\eta^2) U_\eta=0 \; , \hbox{ in
$\Omega_f\setminus\overline{\Theta_\eta}$} \; , \;
\label{6-helmholtz}
\end{equation}
since $D_\eta$ vanishes inside the inclusions $\Theta_\eta=\bigcup_{i\in\mathbb{Z}^2}\{ \eta (i+C) \}$ and it is a constant in the matrix.
Bear in mind that the number of SRR in $\Omega_f$ is an integer which scales as $\eta^{-2}$.
Note also that the vanishing of bending moment and shearing stress deduced from (\ref{bihaeta}) requires that
\begin{equation}
\begin{array}{ll}
\left((1+\nu_\eta)\left(\frac{\partial^2}{\partial x_1^2}+\frac{\partial^2}{\partial x_2^2}\right)
+2(1-\nu_\eta)\left(\frac{\partial^2}{\partial x_1\partial x_2}\right)\right)U_\eta=0 \; ,
\nonumber \\
\left((3-\nu_\eta)\left(\frac{\partial^3}{\partial x_1^3}+\frac{\partial^3}{\partial x_2^3}\right)
+(1+\nu_\eta)\left(\frac{\partial^3}{\partial x_1^2\partial x_2}
+\frac{\partial^3}{\partial x_1\partial x_2^2}\right)\right)U_\eta=0 \; ,
\end{array}
\end{equation}
At the boundary $\partial\Theta_\eta$ of $\Theta_\eta$, which is consistent with our former work on thin perforated plates \cite{farhat-epl}.

\noindent In the present case, perforations are shaped as split ring resonators, and each SRR $C$ can be modeled as
\begin{equation}
C=\{a < \sqrt{x_1^2+x_2^2} < b\} \setminus\overline{\Pi_\eta} \;
\end{equation}
where $a$ and $b$ are functions of variables $x_1,x_2$,
unless the ring is circular and
\begin{equation}
\Pi_\eta = \Bigl \{ (x_1,x_2) \, : \, 0<x_1 < l \, , \, \mid x_2
\mid < \eta h/2 \Bigr \} \; , \label{thindom}
\end{equation}
is a thin ligament of length $l=b-a$ between the ends of the letter $C$,
see Fig. \ref{fig6}(a).

Our aim is to show that the homogenized multi-structured platonic structure within $\Omega_f$ is characterized by an effective density which can take negative values near the fundamental resonance of the SRR. To do this, we need to perform a homogenization of a periodic structure involving resonant elements. The resonances are associated with fast-oscillating fields in thin-bridges of SRR perforations of the plate, and we filter these oscillations by introducing a third scale in the usual two-scale expansion:

\begin{equation}
\begin{array}{lll}
\forall {\bf x} \in \Omega_f, \;\; U_\eta ({\bf x}) &=
\displaystyle{U_0 ({\bf x},{{\bf x}\over\eta},{x_2\over\eta^2}) +
\eta U_1 ({\bf
x}, {{\bf x}\over\eta},{x_2\over\eta^2})} \nonumber \\
& + \displaystyle{\eta^2 U_2 ({\bf x},{{\bf
x}\over\eta},{x_2\over\eta^2})} +...
\end{array}
\end{equation}

\noindent where $U_i : \Omega_f \times Y\times[-h/2,h/2]\longmapsto\mathbb{C}$ is a smooth function of variables $({\bf x},{\bf y},\xi)=(x_1,x_2,y_1,y_2,\xi)$, independent of $\eta$, such that $\forall {\bf x}\in\Omega_f,\; \phi_i({\bf x},\cdot,\xi)$ is $Y$-periodic, and $h$ denotes the thickness of the thin-cut of the SRR.

\noindent The differential operator is rescaled accordingly as $\nabla=\nabla_{\bf x}+\frac{1}{\eta}\nabla_{\bf y}+\frac{1}{\eta^2}\nabla_{\xi}$, so that Eq. (\ref{6-helmholtz}) can be reexpressed as
\begin{equation}
\begin{array}{lll}
&
\displaystyle{\left\{\sqrt{D_\eta}\left(\nabla_{\bf
x}+\frac{1}{\eta}\nabla_{\bf y} +\frac{1}{\eta^2}\nabla_\xi\right)
\cdot \left(\nabla_{\bf x}+\frac{1}{\eta}\nabla_{\bf y} +
\frac{1}{\eta^2}\nabla_{\xi}
\right) +\beta_\eta^2 \right\}} \nonumber \\
&\times\displaystyle{\left\{ \sqrt{D_\eta}\left( \nabla_{\bf
x}+\frac{1}{\eta}\nabla_{\bf y} +\frac{1}{\eta^2}\nabla_\xi\right)
\cdot \left(\nabla_{\bf x}+\frac{1}{\eta}\nabla_{\bf y} +
\frac{1}{\eta^2}\nabla_{\xi}
\right) -\beta_\eta^2 \right\}} \nonumber \\
&\displaystyle{\left\{\sum_{i=0}^\infty \eta^i U_i({\bf
x},\frac{{\bf x}}{\eta},\frac{x_2}{\eta})\right\}} =0 \; ,
\end{array}
\label{helmholtzrescaled}
\end{equation}
with the notations $\nabla_{\bf x}=(\frac{\partial}{\partial x_1},\frac{\partial}{\partial x_2})$, $\nabla_{\bf y}=(\frac{\partial}{\partial y_1},\frac{\partial}{\partial y_2})$ and $\nabla_\xi=(0,\frac{\partial}{\partial\xi})$.

\noindent Collecting terms of same power of $\eta$ in Eq. (\ref{helmholtzrescaled}), we obtain the
homogenized problem \cite{hombook},
\begin{equation}
(\sqrt{D_{hom}}\nabla^2 U_\eta+\beta_0^2{\sqrt{\rho_{hom}}})(\sqrt{D_{hom}}\nabla^2 U_\eta-\beta_0^2{\sqrt{\rho_{hom}}}) U_\eta=0 \; , \hbox{ in
$\Omega_f\setminus\overline{\Theta_\eta}$} \; , \;
\label{bihahom}
\end{equation}
in the limit when $\eta$ tends to zero, in the platonic crystal $\Omega_f$. 
This is the homogenized biharmonic equation with the homogenized
plate rigidity $D_{hom}={(\int_{Y^*}D^{-1}(y_1,y_2)dy_1dy_2)}^{-1}$ 
where $Y^*=Y \setminus \bar{C}$. We point out that the circular
geometry of SRR is essential, if one would take elliptical SRR, then
$D_{hom}$ would be a rank-2 anisotropic tensor.

\noindent The homogenized density $\rho_{hom}$ is given in the form
\begin{equation}
\sqrt{\rho_{hom}}(\beta)=1-\sum_{m=1}^{\infty}
\frac{\beta^2}{\beta^2-{\beta_m}^2} {\Vert V_m
\Vert}_{L^2(0,l)}^2 \; , \label{negdens}
\end{equation}

\noindent where the eigen-solutions $V_m$ correspond to longitudinal vibrations of the thin cut $\Pi_\eta$ within the split ring resonator.

\noindent It is clear from Eq. (\ref{negdens}) that $\sqrt{\rho_{hom}}(\beta)$ takes negative values near resonances $\beta^2=\beta_m^2$. For our purpose, it is enough to look at the first few resonant frequencies $\beta_m^2$ (the higher the frequency the worse the asymptotic approximation). These frequencies are associated with vibrations $V_m$ of the thin domain $\Pi_\eta$ \cite{Seb_SRR}:
\begin{eqnarray}
V_m''(x_1) + {\beta_m}^2 V_m(x_1) & = & 0 \, , \, 0<x_1<l \, , \label{6-sys1}\\
V_m(0) & = & 0 \; , \label{6-sys2}\\
\eta h V_m'(l) & = & \mathrm{area}(\Xi) {\beta_m}^2 V_m(l) \, ,
\label{6-sys3}
\end{eqnarray}
where $\eta h$ and $l$ are the thickness and the length of the thin ligament $\Pi_\eta$, and $\Xi$ is the central disc within the SRR. The ligament $\Pi_\eta$ is connected to $\Xi$, hence $V(l)=V$, where $V$ is the vibration of the stress-free air cavity $\Xi$. Note that the derivation of Eq. (\ref{6-sys3}) required a boundary layer analysis, and we refer to \cite{kozlov99} for more details.

\noindent The solution of the problem (\ref{6-sys1})-(\ref{6-sys3}) has the form
\begin{equation}
V_m(x_1)= A \sin(\beta_m x_1) \; , \label{asymfield}
\end{equation}
where $\beta_m$ (square root of frequency) is given as the solution of the following equation
\begin{equation}
\eta h\cot(\beta_m l) = \mathrm{area}(\Xi) \beta_m \; .
\label{tranres}
\end{equation}

Note also that a similar problem to
Eqs. (\ref{6-sys1})-(\ref{6-sys3}) is deduced from Eq. (\ref{helmholtzrescaled}) with a minus sign in (\ref{6-sys1}), which leads to
a $\sinh$ function in (\ref{tranres}), but this does not give any additional resonant frequencies. The resonant
frequencies of $\sqrt{\rho_{hom}}$ also lead to negative values of $\rho_{hom}$, in the same way that a
product of two simultaneously negative square roots give a negative refractive index in the metamaterial's literature
\cite{veselago,ramak2005}.


In the sequel, we focus our analysis on the Floquet-Bloch bending wave problem and give some numerical results and comparisons with the homogenization theory.

\section{Numerical analysis of SRR platonic crystals}

\subsection{Dispersive properties}

\begin{figure}[h]
\begin{center}
\scalebox{0.65}{\includegraphics[angle=0]{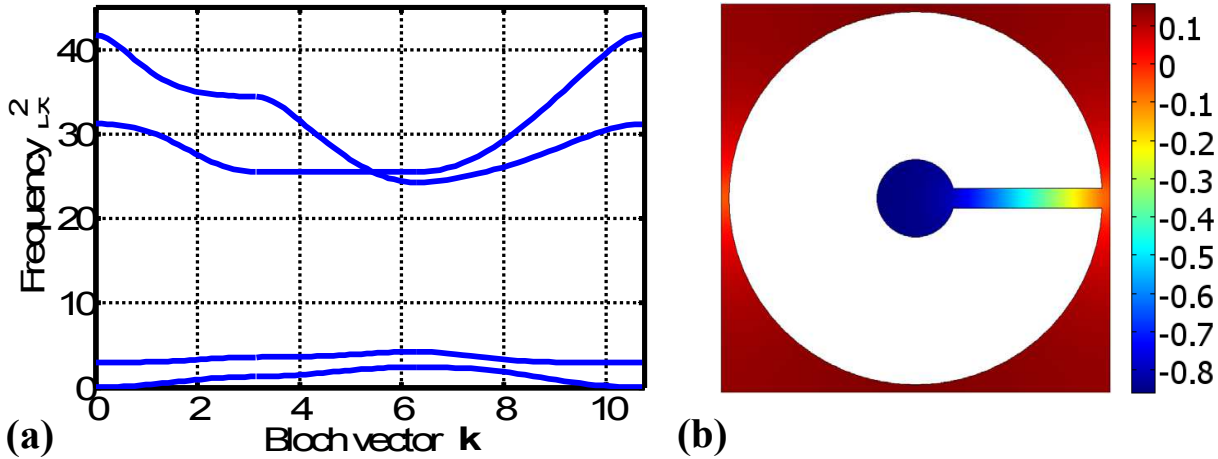}}
\mbox{}\vspace{-5.5cm}\caption{(a) Band diagram for the doubly periodic square lattice representing the normalized frequency $\beta^2$ versus the Bloch vector {\bf k} in the first Brillouin zone $\Gamma$XM. (b) Map of the eigen-mode corresponding to the eigen-frequency $\beta^2=2.95$ in a unit cell for a SRR with an outer radius $0.48d$ and inner radius $0.1d$ and thickness of the thin ligament equal to $0.05d$.}
\label{fig1}
\end{center}
\end{figure}

To investigate numerically the stop-band properties for out-of-plane bending waves propagating within the array of SRR, we use the Finite Element Method (by implementing the weak form of (\ref{6-helmholtz}) in the commercial software COMSOL and its corresponding Floquet-Bloch boundary conditions as well as the PMLs: perfectly matched layers).

In Fig. \ref{fig1}(a), we give the band diagram for normalized eigen-frequencies $\beta^2$ as a function of the projection of the Bloch vector ${\bf k}$ on the first Brillouin zone $\Gamma$XM. We consider a square array of normalized pitch $d$=1 (take for example $d=1$cm for comparisons with feasible experiments) with embedded SRR of inner radius $0.1d$ and outer radius $0.48d$ with a thin cut (ligament) of thickness $0.05d$.\\
This figure displays two full platonic band gaps for the range of normalized frequencies $[2.39, 2.92]$ and $[4.23, 24.32]$. As depicted in a previous study \cite{Seb_SRR}, the low frequency band gap (the first one) is associated with the resonant modes of a single SRR (microscopic structure), while the second one is due to a Bragg scattering phenomenon (macroscopic structure). We also plot in Fig. \ref{fig1}(b) the corresponding localized eigen-function (the first eigen-mode) which corresponds in the context of continuum mechanics, to oscillations of the central region of the SRR as a rigid solid connected by the thin cut $\Pi_{\varepsilon}$ to the fixed rigid region around it \cite{kozlov99}. In our context of bending waves, this can be interpreted as a collective vibration of the plate elements together, up and down in each SRR.

\begin{figure}[]
\begin{center}
\scalebox{0.65}{\includegraphics[angle=0]{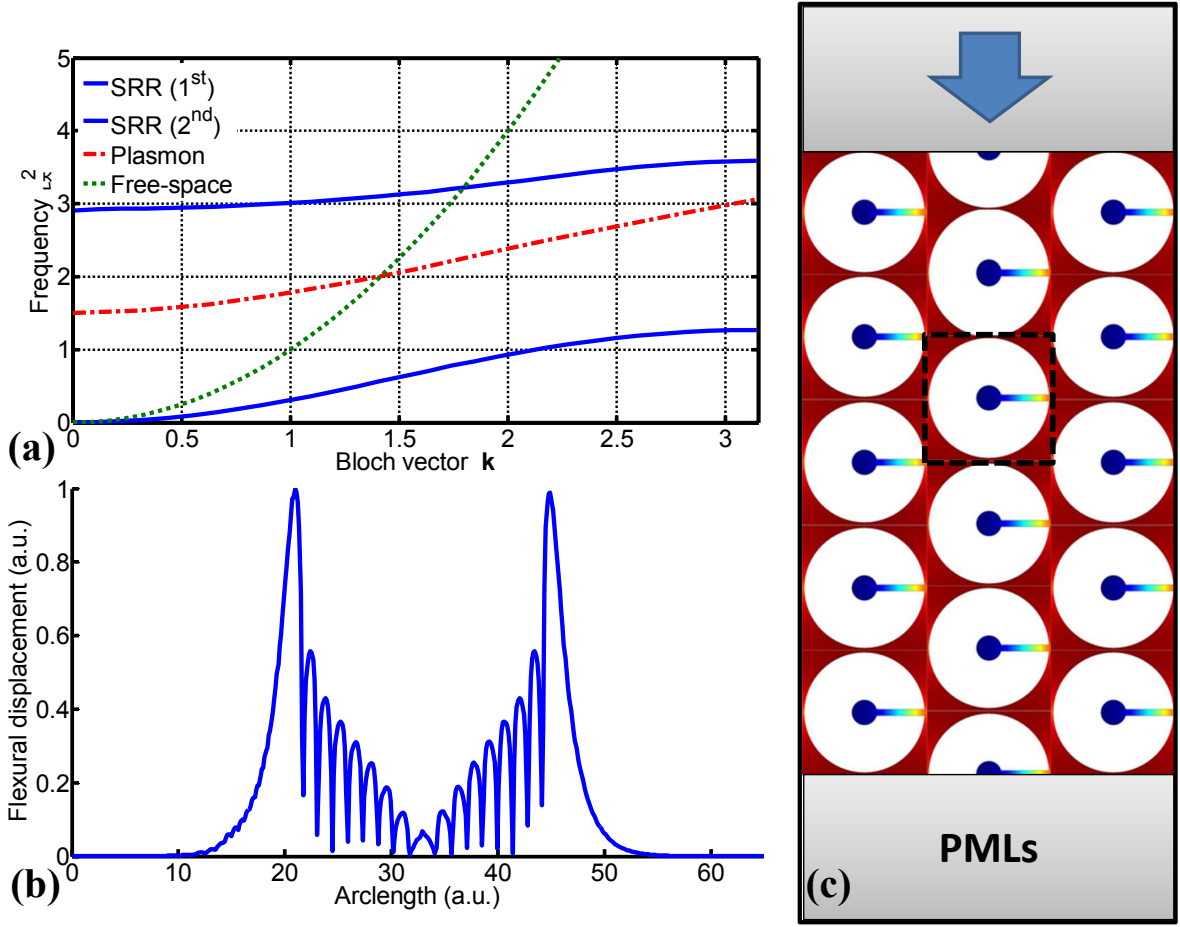}}
\caption{(a) Band diagram of surface mode along the $\Gamma$X direction. (b) 1D profile along the $x_2$ direction of the mode. (c) Supercell showing the configuration where the plasmon mode was excited from the top layer with PMLs in the bottom layer to avoid any reflection and Floquet-Bloch boundary conditions on the left and right sides to account for the infinite array in $x_1$
direction.
}
\label{fig5_2}
\end{center}
\end{figure}

The transcendental equation (\ref{tranres}) can be further simplified if we look at the first low frequency, for which we deduce the explicit asymptotic approximation
\begin{equation}
\beta_1^2 \sim \frac{\eta h}{\mathrm{area}(\Xi)l} \; .
\label{asymfor}
\end{equation}

\noindent Our numerical estimate is
\begin{equation}
\beta_1^2 \sim \frac{0.05}{0.38} \frac{1}{\pi
0.1^2}=4.19 \; ,\label{numestimate1}
\end{equation}
which is in excellent agreement with the finite element value for the plasmon frequency $\beta^2=4.23$ occurring at the $M$ point when ${\bf k}=(\pi,\pi)$.

It is interesting to also analyze the dispersive properties of surface elastic modes (waves exponentially localized at the interface of a platonic crystal). These modes originate from interference effect in phononic or platonic crystals and are important, for example, for superlensing effect. The dispersion diagram of these modes is shown in Fig. \ref{fig5_2}(a), where the frequency is plotted versus Bloch vector in the $\Gamma$M direction. The red dotted-dashed line reprsents the surface mode's dispersion and shows that's standing in the elastic bandgap (between the first and the second propagating modes). This means, for instance, that this mode is of evanescent nature. Figure \ref{fig5_2}(b) shows the flexural displacement at the interface between the platonic crystal and free-space [Fig. \ref{fig5_2}(c)], validating the decaying nature of the mode.

\subsection{(Super-)Lensing effect via negative effective density of a platonic crystal}

\begin{figure}[h]
\begin{center}
\scalebox{0.6}{\includegraphics[angle=0]{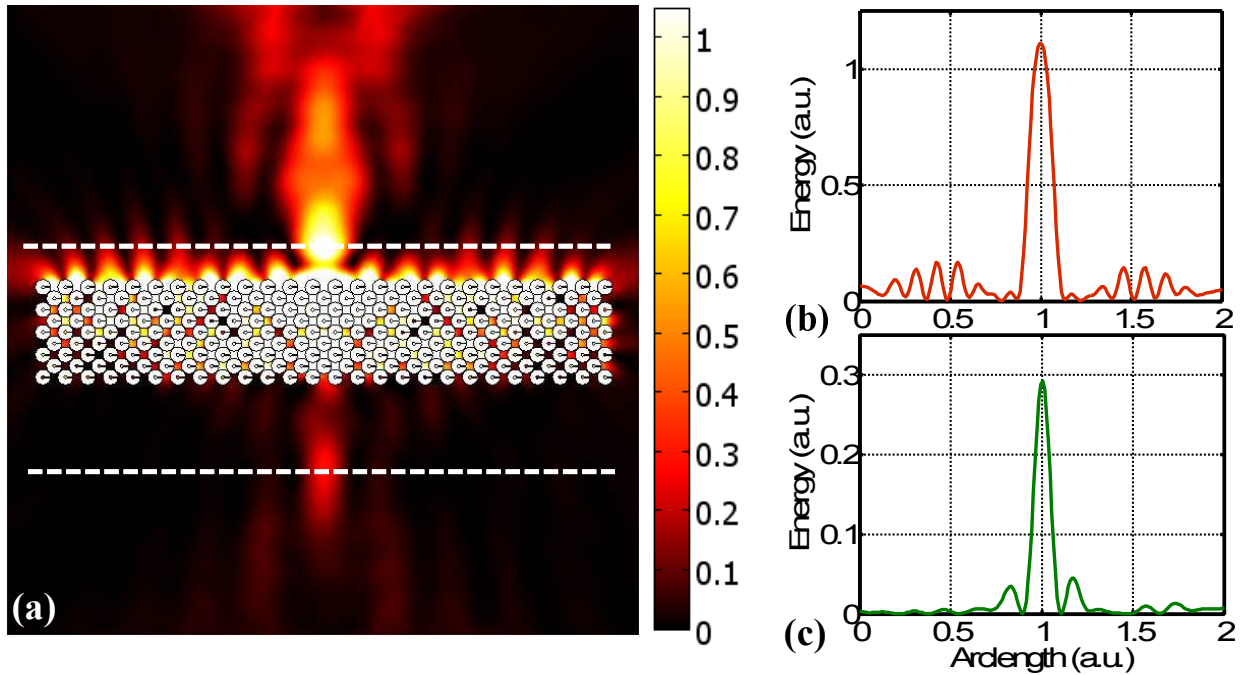}}
\mbox{}\vspace{-3cm}\caption{(a) Pattern of the energy of the bending wave at the frequency $\beta=1.98$ where AANR occurs. (b) and (c) give respectively the profile of the energy along the dotted white line at the source (red curve) and the image (green curve) axis, respectively.}
\label{fig2}
\end{center}
\end{figure}

We then consider a 2D finite platonic crystal perforating a thin elastic plate. The crystal consists of 230 holes with a C-shaped cross-section with identical parameters as those of Fig. \ref{fig1}(b). The crystal has a four-fold (square) symmetry with pitch $d=1$. A flexural waves source of wavenumber $\beta=1.98$ is located at a distance $0.1d$ above the top of the crystal. The field-maps in Fig. \ref{fig2} (a) show the existence of an image of the source, located twice the width of the crystal away in the lower side. As we can see in Fig. \ref{fig2} (a), the amplitude of the field is the highest near the thin ligaments $\Pi_{\varepsilon}$ of the SRR. Formula (\ref{negdens}) shows that the resonance of the field with the microscopic structure of SRR is responsible for the appearance of negative refraction through negative effective density $\rho_{hom}$. Finally, the resolution of the image $\delta$ is enhanced compared to that obtained using an array of perforations with a circular or square cross-section, as demonstrated in Fig. \ref{fig2}(b), (c), whereby the full width at half maximum of the image point $\delta\approx\lambda/3$.

\noindent We note that at the plasmon frequency, the vibration of the thin ligaments is enhanced, which is in accordance with the behavior of the field as observed on Fig. \ref{fig2}. It is therefore important to incorporate the resonances within the thin-bridges as we did, to model the frequency at which negative refraction occurs.

\subsection{Gaussian beam at oblique incidence on a platonic crystal}

\begin{figure}[h]
\begin{center}
\scalebox{0.65}{\includegraphics[angle=0]{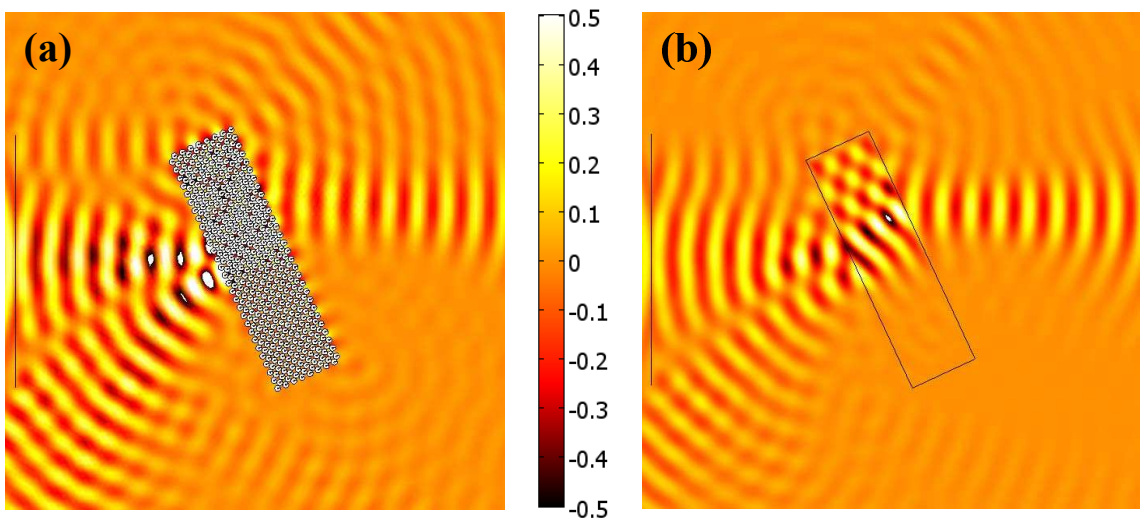}}
\mbox{}\vspace{-4.5cm}\caption{Bending displacement when the platonic crystal is excited by a gaussian beam at an incidence angle $\theta_{\mathrm{inc}}=25$ degrees. (b) Snapshot of the bending wave in the presence of an effective medium with negative elastic index $n_{\mathrm{e}}=-1.5+10^{-3}j$, with $n_{\mathrm{e}}=\sqrt{12\rho(1-\nu^2)/(EH^2)}$.}
\label{fig3}
\end{center}
\end{figure}

The interaction of a Gaussian beam (GB) with the platonic crystal around the AANR is considered in this section. Refraction of GB (whose size is in the order of the wavelength) obliquely incident on the SRR platonic crystal is demonstrated to be negative.

\noindent As known, considering the interaction of a GB with any optical or mechanical media, reveals much of their propagating and scattering properties \cite{gbeam,gbeam2}. These are paraxial solutions of the biharmonic scalar wave  equation $(\Delta^2-\beta^4)W=0$, and thus have the form
\begin{equation}
U(x,y)=u(x,y)e^{-j\beta x}+n.p.
\label{gaussian}
\end{equation}9
with the wavenumber $\beta$, the coefficient $u(x=0,y)$ has the well know Gaussian distribution in the $x-y$ plane. $n.p.$ denotes a non propagating contribution due to the fourth order nature of the biharmonic equation. It could be neglected in the real implementation. The expression of the obtained elastic beam is thus
\begin{equation}
u(x,y)=u_0\frac{\omega_0}{\omega(x)}e^{\left(-\frac{y^2}{\omega^2(x)}-ik\frac{y^2}{2R(x)}+i\zeta(x)\right)}\, ,
\label{gaussian}
\end{equation}
with $\omega(x)=\omega_0\sqrt{1+\left(\frac{x}{x_R}\right)^2}$,
$x_R=\frac{\pi\omega_0^2}{\lambda}$ and $\omega_0$ is called
the waist of the beam. The radius of the curvature is given by
$R(x)=x\left[1+\left(\frac{x_R}{x}\right)^2\right]$ and the
Gouy phase shift is $\zeta(x)=\mathrm{atan}\frac{x}{x_R}$.

\noindent Figure \ref{fig3} gives the snapshot of the
displacement field $U$ in the presence of the platonic crystal
at oblique incidence with angle
$\theta_{\mathrm{inc}}=25$ degrees; a reflected beam could be
distinguished as well as a refracted one which turns out to be
in the negative refraction regime. To further verify this
claim, another simulation using, now, an homogeneous effective
slab of the same size, with negative elastic index $n_{\mathrm{e}}=\omega/\beta^2=\sqrt{12\rho(1-\nu^2)/(EH^2)}$,
is shown in Fig. \ref{fig3}(b). The patterns of the wave are identical in \ref{fig3}(a) and
\ref{fig3}(b) and show that the elastic index of the SRR crystal could be
described as negative: $n_{\mathrm{e}}=-1.5+10^{-3}j$. The
robustness of the AANR was further verified by changing the
angle of incidence $\theta_{\mathrm{inc}}$ and the wavelength
of operation around their main values, and the effect still
could be observed. 

\subsection{Ultrarefraction effect via vanishing effective density of a platonic crystal}

\begin{figure}[h]
\begin{center}
\mbox{}\vspace{-4.5cm}\scalebox{0.65}{\includegraphics[angle=0]{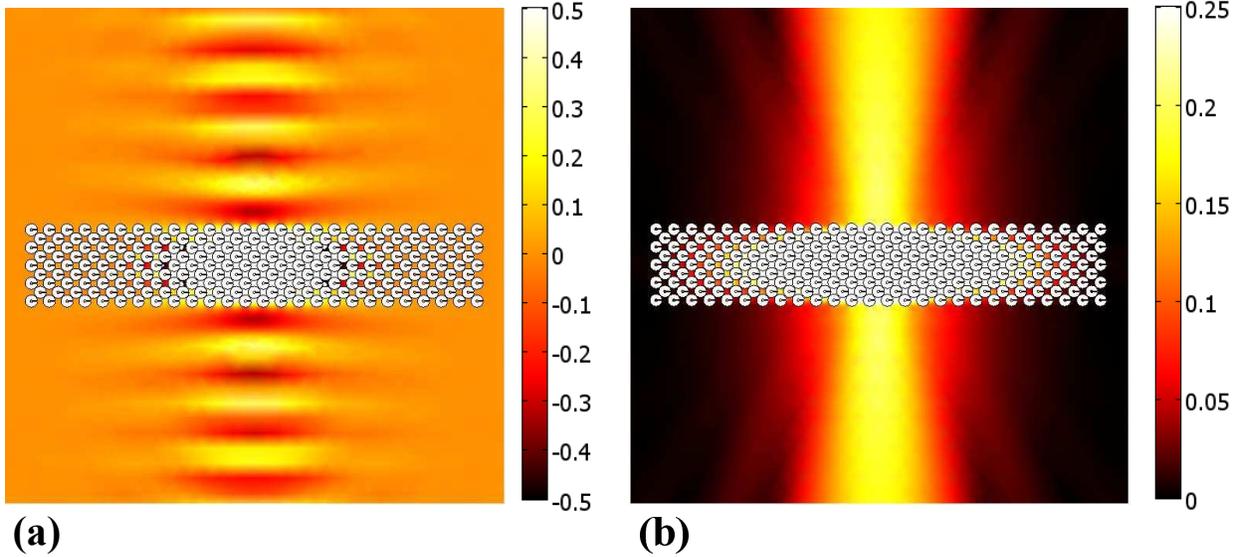}}
\caption{Pattern of the elastic displacement (a) and energy of the bending wave (b) at the frequency where ultrarefraction occurs when a point source is located at the center of the platonic crystal.}
\label{fig4}
\end{center}
\end{figure}

To conclude this section, let us consider another anomalous refractive effect, namely, the ultra-refraction of bending waves. This phenomenon generally occurs when the effective elastic index $n_{\mathrm{e}}$ defined above, tends to zero or equally speaking the group velocity ${\bf v}_{\mathrm{g}}\approx0$. In this case, independently of the angle of incidence of an incident ray, propagating towards the crystal, it will be refracted normally to it (parallel to the normal). And the operation point could be easily deduced from the dispersion diagram by picking a point where the mode's dispersion curve becomes flat ($\nabla_{\bf k}\beta^2\approx0$). This allows us to build an elastic antenna (for instance by placing a cylindrical source of flexural waves in the center of the crystal). The wave will be then refracted along the normal to the crystal, permitting thus highly directive beam as could be seen from Fig. \ref{fig4}(a) and (b) where the displacement snapshot and its energy are respectively plotted around a wavelength $\lambda=2\pi/\beta\approx3$. This corresponds as previously stated to the $M$ point in the diagram of Fig. \ref{fig1}(a) at the maximum. Using formula (\ref{negdens}), one can indeed achieve ultrarefraction when the effective density of the platonic crystal $\rho_{hom}$ is close to zero, which happens for
$\beta_2^2\sim 4.37$. To get this estimate, one need replace the boundary condition (\ref{6-sys2}) by $V_m(0)=-\rm{area}(\Sigma)/\rm{area}(Y-C)$ which amounts to assuming that the field takes the same {\it non zero} values on the opposite edges of the basic cell.
This leads to the new frequency estimate
\begin{equation}
\beta_2^2 \sim \frac{\eta h}{\mathrm{area}(\Xi)l}
\left( 1 + \frac{\rm{area}(\Sigma)}{\rm{area}(Y-C)}
\right) \; ,
\label{asymforbest}
\end{equation}
which is a refinement of (\ref{asymfor}).
The numerical result $\beta_2^2\sim 4.37$ is in good agreement with finite element computations (4.385).


\section{Concluding remarks}

\noindent In conclusion, we have proposed an original route towards elastic negative refraction (superlensing) and ultrarefraction based on excitation of flexural surface modes. To analyze these effects, we have derived the homogenized biharmonic equation using a multi-scale asymptotic approach. We found that the homogenized elastic parameters are described by a scalar flexural plate rigidity (since the perforations have a circular geometry owing to the fact that thin-bridges come unseen in the second scale asymptotics) and a scalar density, the latter taking negative values near thin-bridges resonances (unveiled by the third scale). We then performed numerical computations based on the finite element method which confirmed that the asymptotic model gives accurate predictions on dispersive properties of the elastic Lamb modes. 

We believe that such a micro-structured plate could be manufactured easily, having in mind some potential applications in superlensing or directive elastic antennas. The range of industrial applications is vast, and our proof of concept should foster research efforts in this emerging area of acoustic and seismic metamaterials.

\acknowledgments
S.G. would like to acknowledge a funding of the European Research Council through ERC grant ANAMORPHISM.

\end{document}